\documentclass[twocolumn,nofootinbib,amsmath,prd,aps,superscriptaddress,tightenlines]{revtex4}
\usepackage[hypertex]{hyperref}
\usepackage{graphicx}
\usepackage{epstopdf}
\usepackage{bm}
\usepackage{epsfig}
\usepackage{graphics}
\usepackage{xspace}

\def\OMIT#1{}

\newcommand{\nn}{\nonumber}

\newcommand{\bea}{\begin{eqnarray}}
\newcommand{\eea}{\end{eqnarray}}

\begin{document}
\setlength\baselineskip{17pt}


\preprint{ \vbox{  \hbox{CALT-68-}
}
}

\title{\boldmath
New Physics effects in Higgs Decay to Tau Leptons}

\vspace*{1cm}

\author{Sonny Mantry}
   \affiliation{California Institute of Technology, Pasadena, CA 91125}
   

\author{Michael J. Ramsey-Musolf}
\affiliation{California Institute of Technology, Pasadena, CA 91125}
\affiliation{University of Wisconsin-Madison, Madison, WI 53706}

\author{Michael Trott}
   \affiliation{University of California at San Diego, La Jolla, CA 92093}



\begin{abstract}
  \vspace*{0.3cm}
We study the possible effects of TeV scale new physics (NP) on the rate for Higgs boson decays to charged leptons, focusing on the $\tau^+\tau^-$ channel which can be readily studied at the Large Hadron collider. Using an SU(3)$_C\times$SU(2)$_L\times$U(1)$_Y$ invariant effective theory valid below a NP scale $\Lambda$, we determine all effective operators up to dimension six that could generate appreciable contributions to the decay rate and compute the dependence of the rate on the corresponding operator coefficients. We bound the size of these operator coefficients based on the scale of the $\tau$ mass, naturalness considerations, and experimental constraints on the $\tau$ anomalous magnetic moment. These considerations imply that contributions to the decay rate, when $\Lambda \sim$ TeV, could be comparable to the prediction based on the SM Yukawa interaction. A reliable test of the Higgs mechanism for fermion mass generation via the $h\to \tau^+\tau^-$ channel is possible only after such NP effects are understood and brought under theoretical control.
\end{abstract}

\maketitle

\newpage

\section{Introduction}

The search for the Higgs boson and the study of its properties is a 
primary 
task of the Large Hadron Collider (LHC). Currently, global fits to precision electroweak data 
find the Higgs
mass to be  $84\genfrac{}{}{0pt}{1}{+33}{-24} {\rm GeV} $ with an 
upper bound given by $
m_h < 150 \, {\rm GeV}$ at $95 \% $ CL \cite{PDBook}. LEP 
has also placed a lower
bound of $ m_h>114.4 \, {\rm GeV}$ \cite{Barate:2003sz}.
Assuming the Standard Model (SM) of electroweak
interactions,  the Higgs is expected to be found in early physics 
runs at LHC or
in the near future at the Tevatron.

If a new scalar particle is found at LHC or the Tevatron, studying its self coupling and its couplings to fermions and gauge bosons will be important steps in determining whether or not it 
is the SM Higgs boson. One promising channel for testing the coupling to fermions is $h\to \tau^+ \tau^-$\cite{Rainwater:2007cp}. One can look at Higgs production via Weak Boson Fusion (WBF) which has distinctive signals (see \cite{Rainwater:2007cp} for a general review, see \cite{Plehn:2001nj} for NP effects on WBF) allowing one to eliminate most of the QCD background. In addition, in WBF the Higgs is typically produced with $p_T \sim 100 \, {\rm GeV} \gg m_\tau$ which facilitates a relatively precise invariant mass reconstruction of the $\tau ^+\tau ^-$ pair.
Realistically, at the LHC a measurement of the $\rm h \, \bar{\tau} \, \tau$ coupling for $m_h < 140 \, {\rm GeV}$ 
is expected to be made with about 100 ${\rm fb}^{-1}$ of data  to $\sim 10 \%$ accuracy \cite{Zeppenfeld:2000td,Duhrssen:2004cv}. 

In this paper, we examine how new physics (NP) at or above the $\rm TeV$ scale could effect the decay rate $\Gamma (h\to \tau ^+ \tau ^-)$. If such effects are large, they could complicate a test of the SM Higgs mechanism for fermion mass generation using this decay channel.  Of course identical statements can be made for $h\to e^+ e^-, \mu^+ \mu^-$ although these channels are too  suppressed by small Yukawa couplings to be experimentally interesting (we will briefly comment on these particular channels). In order to analyze possible NP effects on $\Gamma(h\to\tau^+\tau^-)$ in a model-independent manner, we employ an effective field theory approach where NP effects are encoded in $SU(3)\times SU(2) \times U(1)$ invariant higher dimension operators
built out of SM fields:
\begin{equation}
\mathcal{L}_\textrm{eff}=
{\displaystyle\sum\limits_{n,j}}
\frac{C_{j}^{n}(\mu)}{\Lambda^{n-4}}\mathcal{O}_{j}^{(n)}(\mu)\, +\, {\rm h.c.} ,
\label{eq:Leff}
\end{equation}
where $\mu$ is renormalization scale, $n\geq 4$ denotes 
the operator dimension, and $j$ is the index running over all independent operators. 

In what follows, we will take the NP scale $\Lambda$ to be at or above the TeV scale.
New physics at such a  scale is expected 
for at least two reasons: triviality asserts that the Higgs mass 
vanishes in the absence of a cut off \cite{Dashen:1983ts,Kuti:1987nr},
and the radiative instability in the Higgs sector in the absence of additional NP leads to the hierarchy problem. 
Such an EFT approach, with $\Lambda \sim$ TeV, has been applied
to precision electroweak 
observables\cite{Weinberg:1979sa,Wilczek:1979et, Leung:1984ni,
Buchmuller:1985jz,Grinstein:1991cd,Hagiwara:1993ck} and has
recently been the subject of further investigations for LHC and ILC 
phenomenology
\cite{Grinstein:2007iv,Manohar:2006gz,Manohar:2006ga,Kile:2007ts,Fox:2007in,Graesser:2007yj,Hankele:2006ma,Klamke:2007pn,Klamke:2007cu} as well as neutrino properties and interactions \cite{Bell:2005kz,Davidson:2005cs,Bell:2006wi,Erwin:2006uc}.

We use naturalness and/or experimental constraints to bound the Wilson coefficients of the relevant higher dimension operators. 
We find for $h \rightarrow 
\ell^+ \, \ell^-$, where $\ell =\{e,\mu,\tau \}$,  that
$ \Delta \, \Gamma(h \rightarrow \ell^+ \, \ell^-)/\Gamma  \equiv 
(\Gamma_{SM + NP} - \Gamma_{SM})/\Gamma_{SM} \sim
v^2/\Lambda^2 \, \times 1/y_{\ell}\times  \, C$, where $v=246$ GeV is the Higgs vacuum expectation value, $y_\ell$ is the charged lepton Yukawa coupling, and $C$ denotes a combination of  Wilson coefficients of the higher dimension operators. 
Given that $1/y_{\ell} \gg 1$, one might naively expect that very large deviations from the SM rate could be observed. As we show, naturalness considerations generally imply that $C\sim y_\ell$, thereby counteracting the $1/y_\ell$ enhancement. 
Nevertheless, when $\Lambda$ is not too large compared to $v$, we find that $\Delta\Gamma/\Gamma$ can be of order unity. In this case, a reliable test of the Higgs mechanism for lepton mass generation would require additional studies to disentangle the effects of NP in the $h\to\ell^+\ell^-$ channel.

\section{Higher Dimension Operators} \label{sec2}

 The lowest dimension operator that contributes to $h\to \ell^+\ell^-$ is the $n=4$ SM Yukawa interaction
\begin{equation}
y_\ell \mathcal{O}_{eY} +\textrm{h.c.}\ , \quad \mathcal{O}_{eY} \equiv {\bar L} \phi e\\
\end{equation}
where $L$ and $e$ are lepton SU(2)$_L$ doublet and right handed charged lepton singlet fields respectively and $\phi$ is the Higgs doublet\footnote{We work in a basis where the charged lepton Yukawa matrices are diagonal.}. 
The effects of new physics first appear at $n=6$. Since $\Lambda \sim \> $TeV$\gg v$,   contributions from $n>6$ operators can be safely omitted. Using the basis of 
Buchmuller and Wyler \cite{Buchmuller:1985jz}, the operators relevant to the $h \, \ell^+ \, \ell^-$ coupling at tree level are
\bea
\mathcal{O}_{eH} & = & ( \phi^\dagger  \phi ) {\bar L}\phi e, \nn \\
\mathcal{O}_{H\ell}^{(1)} & = & i( \phi^\dagger D_\mu 
\phi)({\bar L} \gamma^\mu L), \nn \\
\label{eq:oplist}
\mathcal{O}_{H\ell}^{(3)} & = & i( \phi^\dagger D_\mu 
\tau ^I \,  \phi)({\bar{L}} \gamma^\mu \tau ^I {L}), \\
\mathcal{O}_{He} & = &  i( \phi^\dagger D_\mu 
\phi)({\bar{e}} \, \gamma^\mu e), \nn \\
\mathcal{O}_{De} & = & {\bar{L}} \, (D^\mu  \phi) D_\mu \,  e, \nn \\
\mathcal{O}_{{\bar D}e} & = & {\bar{L}}\overleftarrow{D}_\mu^\dag (D^\mu  \phi) \, e.\nn
\eea
It is also useful to consider the symmetric and antisymmetric combinations of the last two operators in Eq.~(\ref{eq:oplist})
\begin{equation}
\mathcal{ O}_{\pm}  =  {\bar{L}}(D^\mu  \phi) D_\mu \,  e \pm {\bar{L}}\overleftarrow{D}_\mu^\dag (D^\mu  \phi) \, e \ \ \ .
\end{equation}

It is straightforward to show that the symmetric combination $\mathcal{O}_+$ can be expressed as  linear combination of  $\mathcal{O}_{eY}$, $\mathcal{O}_{eH}$, and four fermion operators that do not contribute to $\Gamma(h \to \ell^+\ell^-)$ at tree level. Before electroweak symmetry breaking (EWSB), we have 
\bea
\label{eq:totalderiv}
\partial^\mu \left[ {\bar L} (D_\mu \phi) e\right] &=& m^2 \mathcal{O}_{eY}-\lambda\mathcal{O}_{eH}+\mathcal{O}_+ \\
&-& y_e^{\dagger} \, ({\bar L} \, e) \, ({\bar e} \, L) - y_u \, ({\bar L}_b \, e) \, ({\bar q}_a \, \epsilon_{ab} \,  u) \nn\\
&-&  y_d^\dagger \, ({\bar L} \, e) \, ({\bar d} \, q),\nn
\eea
where the current ${\bar L} (D_\mu \phi) e$ is gauge invariant and non-anomalous and the
scalar potential $V_H(\phi)$ is given by 
\begin{equation}
V_{H}(\phi)=-m^2\, \phi^\dag\phi+\frac{\lambda}{2}\, \left(\phi^\dag\phi\right)^2\ \ \ .
\end{equation}
The total derivative on the LHS of Eq.~(\ref{eq:totalderiv}) yields a vanishing contribution to the action associated with $\mathcal{L}_\textrm{eff}$, so one can eliminate $\mathcal{O}_+$ in terms of the other operators appearing on the RHS. For purposes of this analysis, it is convenient to eliminate one of the four-fermion operators apearing in Eq.~(\ref{eq:totalderiv}) and to retain both $\mathcal{O}_{eH}$ and $\mathcal{O}_+$ explicitly.

\section{$n=6$ contributions to $h \to \ell ^+ \ell ^-$ Decay}

Contributions to $\Gamma(h\to \ell^+\ell^-)$ can be obtained by expanding
the Higgs  field as usual around its vacuum expectation value
$v$, so that 
\begin{eqnarray}
\phi (x) = \frac{{\rm  U(x)}}{\sqrt{2}} \,
\left(
  \begin{array}{c}
  0 \\
v + h(x)
\end{array}
\right).
\end{eqnarray}
Here $ {\rm U(x)} = e^{i \, \xi^a(x) \, \sigma_a/v}$ and $\xi ^a(x)$ are the Goldstone boson fields. In unitary 
gauge ${\rm U(x)} = 1$ and all couplings to the Goldstone bosons vanish.  In general, we consider only contributions  to $\Delta\Gamma$ that are linear in the $n=6$ operator coefficients and that come with one power of $v^2/\Lambda^2$. Such terms can only arise from the interference of the SM Yukawa amplitude for the decay with the corresponding amplitude generated by one of the $n=6$ operators\footnote{In some instances, however, the naturalness considerations discussed below imply NP contributions to $\Delta\Gamma$ of the same size as the SM rate, and we retain the quadratic terms in these cases.} . 

\OMIT{Contributions of $\mathcal{O}(v^4/\Lambda^4)$  from the $n=6$ operators would depend quadratically on the corresponding operator coefficients and are the same size as the effects of $n=8$ operators that interfere with the Yukawa amplitude. }

Because the operators $\mathcal{O}_{H\ell}^{(1,3)}$ and $\mathcal{O}_{He}$ contain only same chirality lepton fields, they cannot interfere with $\mathcal{O}_{eY}$ 
. Consequently, their contributions will 
appear at $\mathcal{O}(v^4/\Lambda^4)$, as we have verified by explicit computation. The $\mathcal{O}(v^2/\Lambda^2)$ effects, therefore, are generated by $\mathcal{O}_{eH}$, $\mathcal{O}_{De}$, and $\mathcal{O}_{{\bar D}e}$,  whose amplitudes interfere with $\mathcal{O}_{eY}$ 
. We also observe that only the symmetric combination $\mathcal{O}_+$ contributes at $\mathcal{O}(v^2/\Lambda^2)$ since $\mathcal{O}_{-}$ can be written as a linear combination of the magnetic moment operators that do not contribute to the decay rate and the $n=6$ operators whose effects are chirally suppressed (see below). Thus, we need to consider only the effects of two operators: $\mathcal{O}_{eH}$ and $\mathcal{O}_+$. 

Computing the shifts $\Delta\Gamma$ from these two operators is a straightforward exercise. The relevant Feynman rules  for $\mathcal{O}_{eH}$ are identical to those for ${\cal O}_{eY}$ up to the overall difference in operator coefficients. For $\mathcal{O}_+$, we exploit Eq.~(\ref{eq:totalderiv}) to relate it to $\mathcal{O}_{eY}$ and $\mathcal{O}_{eH}$ and the four fermion operators that do not contribute to $\Delta\Gamma$ at tree-level. The corresponding Feynman rules for $\mathcal{O}_+$ can then be obtained by the appropriate combination of those for the Yukawa interaction and $\mathcal{O}_{eH}$. Thus, we have 
\bea \label{eq:frs}
y_\ell \, \mathcal{O}_{eY} & \rightarrow & \frac{y_\ell}{\sqrt{2}}\, {\bar \ell} P_R \ell \, h, \nn \\
\frac{C_{eH}}{\Lambda^2}\, \mathcal{O}_{eH} & \rightarrow & \frac{3 C_{eH}v^2}{2\sqrt{2}\Lambda^2} {\bar \ell} P_R \ell\, h,\\
\frac{C_+}{\Lambda^2}\, \mathcal{O}_+ & \rightarrow & \frac {C_+ }{2\sqrt{2}\Lambda^2} \left(3\lambda^2v^2-2 m^2\right)\,  {\bar \ell} P_R \ell\, h , \nn \\
&=&\frac {C_+ m_h^2}{\sqrt{2}\Lambda^2} {\bar \ell}\, P_R \ell \, h,\nn
\eea
where $\ell$ is the charged lepton field, $P_R$ is the right handed projection operator and where, in the obtaining the last line of Eq.~(\ref{eq:frs}), we have used the conditions for EWSB to relate $m^2$ and $\lambda v^2$ to the Higgs mass squared $m_h^2$. 

The contribution to $\Delta\Gamma$ from $\mathcal{O}_{eH}$ and $\mathcal{O}_+$ can be read off from Eq.~(\ref{eq:frs}) by using the SM rate
\begin{equation}
\Gamma(h\to \ell^+\ell^-)_\textrm{SM} =\frac{y_\ell^2\, m_h}{16\pi} \, \left(1-4m_\ell^2/m_h^2\right)^{3/2} \ \ ,
\end{equation}
and replacing  $y_\ell$ by ${\bar y}_\ell +3 \,  C_{eH} v^2/2\Lambda^2 + \, C_+ m_h^2/\Lambda^2$, where ${\bar y}_\ell$ denotes the coefficient of ${\cal O}_{eY}$ in the presence of NP. In general, the appearance of a higher-dimension operator that contributes to the lepton mass will change the relationship between the Yukawa coupling and $m_\ell$, implying that ${\bar y}_\ell\not= y_\ell$. In the SM, this relationship is 
\bea
\label{eq:smyukawa}
y_\ell = \sqrt{2} \, \frac{m_\ell}{v}.
\eea
For the $n=6$ operators considered here, $\mathcal{O}_{eH}$ generates a tree-level contribution to $m_\ell$. In this case,   Eq.~(\ref{eq:smyukawa}) no longer gives the relationship between the lepton mass that appears in the Lagrangian density after EWSB and the coefficient of $\mathcal{O}_{eY}$, and we must replace it by\footnote{We thank Mark Wise for pointing out the need to include this correction.}
\begin{equation}
\label{eq:npyukawa}
y_\ell\rightarrow {\bar y}_\ell = y_\ell+\delta y_\ell\ \  ,
\end{equation} 
where $y_\ell$ is given by its SM value as in Eq.~(\ref{eq:smyukawa}) and $\delta y_\ell$ gives the shift due to the presence of a non-vanishing $C_{eH}$. In contrast, the $\mathcal{O}_+$ contributes to $m_\ell$ only at the one-loop level through its mixing with $\mathcal{O}_{eH}$ and the effects of matching onto $\mathcal{O}_{eY}$ at the scale $\Lambda$ (see below). In this case, the mixing of $\mathcal{O}_+$ with $\mathcal{O}_{eH}$ implies a non-vanishing $\delta y$. 

The resulting expression for the relative change in the decay rate is  
\bea 
\label{NPrat}
\frac{\Delta \, \Gamma (h\to \ell^+ \ell^-)}{\Gamma} 
  &=& 
\frac{(y_\ell+\delta y_\ell+ a_1 v^2/\Lambda^2)^2}{y_\ell^2}-1 \ \ \ ,
\eea
where
\begin{equation}
\label{eq:a1}
a_1 = \left[ \frac{3}{2} \>C_{eH} + \frac{m_h^2}{v^2} \>
C_+ \right]\ \ \ ,
\end{equation}
where we have taken all operator coefficients to be real for purposes of this analysis.

 To the extent that the effects of  $\mathcal{O}_{eH}$ and $\mathcal{O}_+$ on $m_\ell$ and $\Delta\Gamma$ are suppressed by $v^2/\Lambda^2$, we may write Eq.~(\ref{NPrat}) as
\begin{equation}
\label{eq:NPRatapprox}
\frac{\Delta \, \Gamma (h\to \ell^+ \ell^-)}{\Gamma} = \frac{2\left(\delta y_\ell+a_1 v^2/\Lambda^2\right)}{y_\ell}+\mathcal{O}(\frac{v^4}{\Lambda^4}) \ \ \ ,
\end{equation}
 and we have used $C_+=(C_{De}+C_{{\bar D}e})/2$. We have cross-checked the result in Eqs.~(\ref{eq:a1},\ref{eq:NPRatapprox}) by using the operators $\mathcal{O}_{De}$ and $\mathcal{O}_{{\bar D}e}$ directly without employing the equations of motion while noting that $C_+=(C_{De}+C_{{\bar D}e})/2$. We observe that the contribution from $\mathcal{O}_{eH}$ depends on $v^2/\Lambda^2$ while the effect of $\mathcal{O}_+$ varies as $m_h^2/\Lambda^2$. 
For the $h\to \tau^+ \tau^-$ channel we obtain
\bea
\frac{\Delta \, \Gamma (h\to \tau^+ \tau^-)}{\Gamma} &\approx & 200 \times\left(\delta y_\tau+ a_1\,
 \frac{ v^2}{{\Lambda}^2} \right),
\eea
indicating the possibility of significant NP effects for $\Lambda \sim$ TeV and reasonable choices for the Wilson coefficient. We will explore bounds on the size of the Wilson coefficients in later sections and show that naturalness considerations imply that they are generally proportional to $y_\ell$.

\section{Estimates of NP effects on $h \rightarrow \ell^+ \, \ell^-$ }
\label{Bounds}
The expressions in Eq.~(\ref{NPrat},\ref{eq:NPRatapprox}) allow us to estimate the size of possible new physics contributions to the $h\to \ell^+\ell^-$ rate. As discussed in Ref.~\cite{Arzt:1994gp}, the operators $\mathcal{O}_{eH}$, $\mathcal{O}_{De}$, and $\mathcal{O}_{{\bar D}e}$ could be generated by tree level effects of new physics above the scale $\Lambda$. As a result, the corresponding operator coefficients could in principle be ${\mathcal{O}}(1)$  rather than  
$\mathcal{O}(1/16 \pi^2$) as in a naive application of naive dimensional analysis (NDA)\cite{Manohar:1983md}. 
Setting $ v^2/({\Lambda}^2 \, y_\ell) = 1$ one finds that NP can have an $\mathcal{O}(1)$
effect for  
$ v \ll {\Lambda} \lesssim 3 \, {\rm TeV}$ for  $\ell=\tau$ 
,  $ v \ll {\Lambda} \lesssim 12 \, {\rm TeV}$ for  $\ell=\mu$ 
and  $v \ll {\Lambda} \lesssim 170 \, {\rm TeV}$ for $\ell=e$.

The resulting shifts $\Delta\Gamma/\Gamma$ are quite large unless $\Lambda$ is very large compared to $v$ and $m_h$. 
However, because these operators have the same chiral structure as $\mathcal{O}_{eY}$, their  coefficients are likely to be constrained by the scale of the charged lepton mass 
in the absence of large cancellations between the effects of different operators. 
In this section we discuss these \lq\lq naturalness" constraints as well as those following from other chirality changing operators such as the charged lepton magnetic moment.
We show that 
 $\mathcal{O}(1)$ corrections to the SM decay rate are still possible.

We will determine the naturalness expectations for the magnitudes of $C_{eH}(\Lambda)$ and $C_{+}(\Lambda)$ separately. In any specific NP scenario, the values of these coefficients at the scale $\Lambda$ are determined after integrating out the heavy degrees of freedom. The resulting terms in $\mathcal{L}_\textrm{eff}$ -- $C_{eH}(\Lambda)\mathcal{O}_{eH}/\Lambda^2$ and  $C_{+}(\Lambda)\mathcal{O}_{+}/\Lambda^2$ -- will generate contributions to the charged lepton mass after evolving to the appropriate low scale. We will assume that the magnitudes of these contributions are no larger than the magnitude of the charged lepton mass itself. 
\OMIT{When the naturalness bounds are saturated, the NP contribution to $m_\ell$ has the same magnitude as the 
mass itself}
\OMIT{and one has $|\delta y_\ell |=y_\ell$ due to the operator $O_{eH}$} To determine the corresponding contributions to $\Delta\Gamma/\Gamma$, we use the full expression in Eq.~(\ref{NPrat}) since the $\mathcal{O}(\delta y_\ell^2)$ terms [and in some cases, the $\mathcal{O}(a_1^2)$ terms] are not necessarily negligible. We also analyze the effects for both choices for the phase of $C_{eH}$  and $C_+$ and generally quote the most restrictive result as the corresponding naturalness expectation.
\OMIT{ For $O_+$ for the sake of simplicity we consider the case where $O_{eH}$ vanishes and use Eq.~(14) with $\delta y_\ell =0$.}

In deriving the resulting naturalness expectations for the $n=6$ operator coefficients, we will follow the approach used recently in Refs.~\cite{Bell:2005kz,Davidson:2005cs,Bell:2006wi,Erwin:2006uc} to derive constraints on operators relevant to neutrino properties and low-energy weak decays based on the scale of neutrino mass. In doing so, we consider three ways in which the presence of a non-zero $C_{eH}(\Lambda)$ or $C_{+}(\Lambda)$ at the high scale can contribute to $m_\ell$. 
\begin{itemize}
\item [(i)] Through tree-level contributions of the corresponding operator.
\item [(ii)] Through one-loop contributions to the $n=4$ lepton mass operator, $\mathcal{O}_{eY}$ at the high scale.
\item [(iii)] Via one-loop mixing of the $n=6$ operators that depend on momenta between $\Lambda$ and the electroweak scale. Below the EW scale our effective theory must be matched onto a different effective theory in which the $W^\pm$ and $Z$ have been integrated out. The constraints that follow from the latter low energy effective theory are too weak to be interesting. 
\end{itemize}

Although the naturalness  considerations can be applied to constrain $h\to \ell^+ \ell^-$ for $\ell =\{e,\mu ,\tau\}$, we will focus on the phenomenologically more interesting $h\to \tau^+ \tau^-$ channel.

\subsection{Naturalness constraints on $\mathcal{O}_{eH}$}

The analysis of naturalness considerations for $\mathcal{O}_{eH}$ is particularly straightforward, as it generates a tree-level contribution to $m_\tau$
\begin{equation}
\delta m_\tau[\mathcal{O}_{eH}] = \frac{C_{eH}(\Lambda)}{2\sqrt{2}}\, \left(\frac{v}{\Lambda}\right)^2\, v\ \ \ .
\end{equation}
Here, we have omitted corrections associated with the running of $C_{eH}(\mu)$ from the scale $\Lambda$ to $v$ as they are loop and coupling suppressed and do not substantially affect the corresponding naturalness expectation\footnote{Recall that we are taking $C_{eH}$ to be real in this analysis.}. 
In the absence of large cancellations between this contribution and SM Yukawa contribution, we have $|\delta m_\tau | \lesssim m_\tau$ or
\begin{equation}
\label{eq:Cehtree}
\left| C_{eH}(\Lambda) \right| \, \frac{v^2}{\Lambda^2}  \lesssim 2 \,  y_\tau \ \ \ .
\end{equation}
The resulting shift in the Yukawa coupling is $\delta y_\tau = - y_\tau$ ($y_\tau$) for positive (negative) $C_{eH}$. Using Eq.~(\ref{NPrat}) we obtain the naturalness bounds $\Delta\Gamma/\Gamma=8$ ($\Delta\Gamma/\Gamma=0$). It is interesting to note that these bounds are lepton species-independent since the RHS of Eq.~(\ref{eq:Cehtree}) is proportional to the Yukawa factor, thereby canceling the factor of $y_\tau^2$ in the denominator of Eq.~(\ref{NPrat}).

When $\Lambda \gg v$, the upper bound on $|C_{eH}|v^2/\Lambda^2$  Eq.~(\ref{eq:Cehtree}) can allow the magnitude of the operator coefficient to be much larger than unity. In addition to appearing physically unreasonable, the presence of very large effective operator coefficients invalidates the truncation of the expansion in Eq.~(\ref{eq:Leff}) at any order. Consequently, we do not consider Eq.~(\ref{eq:Cehtree}) to be physically meaningful for $\Lambda \gg v$. 
A more stringent expectation for the possible magnitude of $C_{eH}$ can be derived  for large $\Lambda$ (see below) by considering the effects of $\mathcal{O}_{eH}$ at one-loop order.  
Effects of loop momenta of order $\Lambda$ can also generate contributions from $\mathcal{O}_{eH}$ to the 
$n=4$ Yukawa interaction $\mathcal{O}_Y$ as in Fig.1 

\begin{figure}[hbt]
\centerline{\scalebox{1.0}{\includegraphics{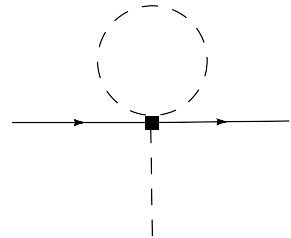}}}
\caption{The one loop contribution of $\mathcal{O}_{eH}$ to $\mathcal{O}_Y$.}
\end{figure}

These contributions will appear when the full theory above the scale $\Lambda$ is used to compute renormalization of $\mathcal{O}_{eY}$ and, thus, would generate a matching correction to the effective theory below the scale $\Lambda$. Without knowing the full theory, we cannot compute this matching contribution precisely. Nevertheless, it is possible to estimate its magnitude using NDA. Doing so yields
\begin{equation}
\delta m_\tau[\mathcal{O}_{eH}\to\mathcal{O}_Y]\sim \frac{C_{eH}(\Lambda)}{16  \, \pi^2}\, \frac{3 \,v}{\sqrt{2}} \ \ \ ,
\end{equation}
leading to
\begin{equation}
\label{eq:Cehloop}
\left| C_{eH}(\Lambda) \right|   \lesssim \frac{16 \, \pi^2}{3} \, y_\tau \, \,  \ \ \ .
\end{equation}
The resulting expectation for the possible size of $\Delta \Gamma/\Gamma$ becomes more  stringent than the tree-level bound when $\Lambda\gtrsim 4\pi v$ since the corresponding contribution to the RHS of Eq.~(\ref{NPrat}) decreases as $v^2/\Lambda^2$. 

It is possible that details of a specific model for NP above the scale $\Lambda$  will preclude any contributions from $\mathcal{O}_{eH}$ to $\mathcal{O}_{eY}$, in which case the naturalness expectation in Eq.~(\ref{eq:Cehloop}) would not apply. In the absence of such a specific scenario, however, Eq.~(\ref{eq:Cehloop}) gives a reasonable estimate of the magnitude of $C_{eH}(\Lambda)$.  

We illustrate the expectations for $\Delta\Gamma/\Gamma$ obtained from Eqs.~(\ref{eq:Cehtree}) and (\ref{eq:Cehloop}) in Fig. 2. \footnote{The effects of the $O_{eH}$ operator on higgs Yukawa couplings has been recently studied within the context of multi-scalar doublet models in ~\cite{twodoublet}. Large regions of parameter space were found where order one effects are realized in the higgs decay rate in agreement with the large effects found to be possible in our naturalness bounds.}


\begin{figure}[hbt]
\centerline{\scalebox{1.1}{\includegraphics{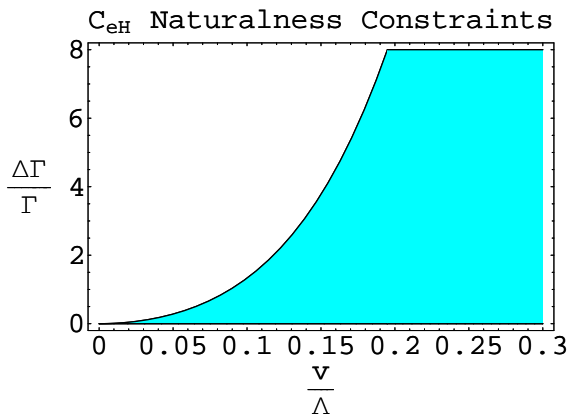}}}
\caption{Combined tree and loop level naturalness bounds on $\Delta\Gamma/\Gamma$ arising from $\mathcal{O}_{eH}$. The region between the curve and the x axis is
the space of allowed $\Delta\Gamma/\Gamma$ due to naturalness constraints. The result shown is for $\delta_y = - y$.}
\end{figure}


Under renormalization, $\mathcal{O}_{eH}$ will mix with other $n=6$ operators, including $\mathcal{O}_{De}$ and $\mathcal{O}_{{\bar D}e}$ as in Fig.4. 
This mixing among like-dimension operators is sensitive to loop momenta from the scale $\Lambda$ to the scale $\mu$ associated with the physical process. 
In the case of $\mathcal{O}_{eH}$, these one-loop mixing effects are dominated by operator self-renormalization, yielding a small correction to the tree-level bound in Eq.~(\ref{eq:Cehtree}). 

\subsection{Naturalness constraints on $\mathcal{O}_{+}$}
We first observe that $\mathcal{O}_{+}$ does not contribute to $m_\tau$ at tree-level since it contains a covariant derivative 
acting on $\phi$. Alternately, we can express Eq.~(\ref{eq:totalderiv}) as
\bea
\label{eq:totalderivalt}
\partial^\mu \left[ {\bar L} (D_\mu \phi) e\right] &=& -{\bar L}\, \left(\frac{\delta V_H}{\delta\phi}\right)\, e +\mathcal{O}_+ \\
&-& y_e^{\dagger} \, ({\bar L} \, e) \, ({\bar e} \, L) - y_u \, ({\bar L}_b \, e) \, ({\bar q}_a \, \epsilon_{ab} \,  u) \nn\\
&-&  y_d^\dagger \, ({\bar L} \, e) \, ({\bar d} \, q),\nn
\eea
so that $\mathcal{O}_+$ can be expressed as a linear combination of four fermion operators that do not contribute to $m_\tau$ at tree level and
\begin{equation}
{\bar L}\, \left(\frac{\delta V_H}{\delta\phi}\right)\, e\ \ \ .
\end{equation}
Since the condition for EWSB is given by $\delta V_H/\delta\phi=0$, $\mathcal{O}_+$ does not contribute to $m_\tau$ at tree-level.

At one loop order,  $\mathcal{O}_+$  generates contributions to both $\mathcal{O}_{eH}$ and $O_{eY}$. Illustrative, one-loop matching contributions are shown in Fig. 3. 
\begin{figure}[hbt]
\centerline{\scalebox{1.1}{\includegraphics{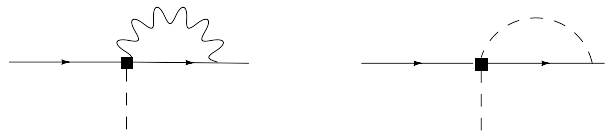}}}
\caption{Two representative diagrams for the one loop matching of $O_{+}$ onto $\mathcal{O}_{eY}$.}
\end{figure}
The EW loops give naturalness constraints proportional to $y_\tau$ while the Higgs loops give 
far weaker constraints proportional to $1/y_\tau$.  The strongest constraints are dictated by the strongest effective coupling in the EW sector. 
Since we cannot determine the precise numerical coefficient  on the RHS of Eq.~(\ref{eq:oplustomtau}) we have estimated the matching contribution 
using only the SU(2)$_L$ gauge loops, neglecting the U(1)$_Y$ and Yukawa-suppressed Higgs loop effects. Since we are only interested 
in order-of-magnitude expectations, this is sufficient. We again use NDA and obtain 

\begin{equation}
\label{eq:oplustomtau}
\delta m_\tau[\mathcal{O}_{+}\to\mathcal{O}_{eY}]\sim \frac{C_{+}(\Lambda) g_2^2}{16  \, \pi^2}\, \frac{ \,v}{\sqrt{2}} \ \ \ ,
\end{equation}
leading to
\begin{equation}
\label{eq:Cplusloop}
\left|C_{+}(\Lambda) \right|    \lesssim \frac{4  \, \pi\sin^2\theta_W}{\alpha} \, y_\tau \, \,  \ \ \ .
\end{equation}

The mixing with $\mathcal{O}_{eH}$
is given to lowest order in the lepton Yukawa coupling by the diagrams in Fig 4.
\begin{figure}[hbt]
\centerline{\scalebox{0.9}{\includegraphics{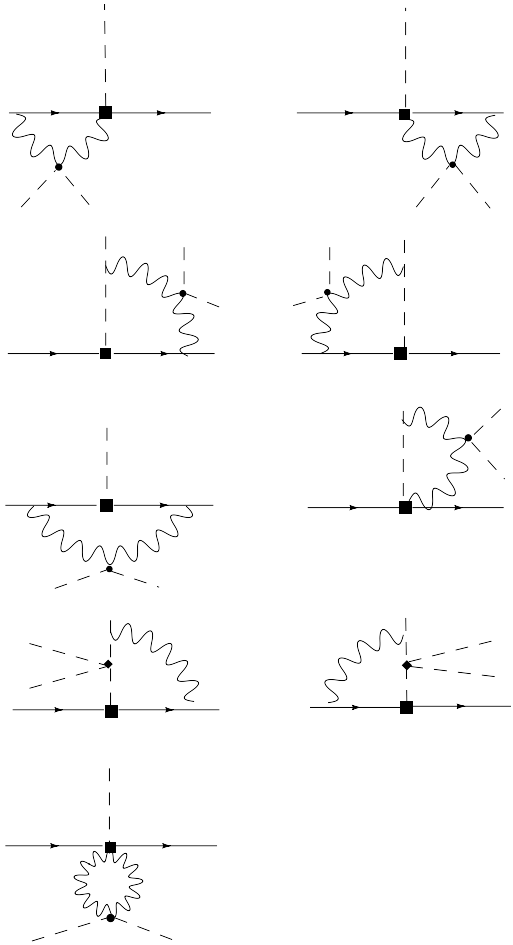}}}
\caption{The one loop contribution of $O_{+}$ to $\mathcal{O}_{eH}$. The dashed lines are $\phi$ fields,
the writhing lines are the SU(2) and U(1) gauge fields. Note that we are above the EW scale v.}
\label{mixing}
\end{figure}
In principle, one can obtain this mixing by computing the anomalous dimension matrix for $\mathcal{O}_{eH}$, $\mathcal{O}_{+}$, and any other independent  $n=6$ operators that mix under renormalization and solving the corresponding renormalization group equations. For the purposes of deriving order-of-magnitude naturalness constraints it suffices to keep only the leading logarithmic contributions (see, {\em e.g.}, Refs.~\cite{Bell:2005kz,Davidson:2005cs,Bell:2006wi,Erwin:2006uc}) which we have computed. 
The contribution to the Wilson coefficient of $O_{eH}$ from $O_+$ via mixing is given by
\begin{equation}
C_{eH}(\mu) = C_{+}(\Lambda)\, {\tilde\gamma}\, \ln\frac{\Lambda}{\mu}+\cdots
\end{equation}
and the dots above denote contributions from self renormalization and the mixing of other operators into $O_{eH}$. We obtain $\tilde{\gamma}$ from the one loop computation of the diagrams in Fig.(\ref{mixing}), we find
\bea
\tilde{\gamma}  &=& \frac{-1}{16 \, \pi^2} \,  \, \left( \frac{3 \,  g_1^4}{8} + \frac{9 \, g_2^4}{8}  +  \frac{\lambda }{2} \,  g_1^2 +   \lambda  g_2^2 + \frac{3}{4} g_1^2 g_2^2\right).\nn \\
\eea
Requiring that the resulting contribution to the $\tau$ mass  be of the same order of magnitude as, or smaller than,  $m_\tau$ leads to the constraint
\begin{equation}
\label{eq:oplusmixbound}
\left\vert C_+(\Lambda)\right\vert \, \frac{v^2}{\Lambda^2} \lesssim \frac{2 y_\tau}{\tilde\gamma}\, \left(\ln\frac{\Lambda}{v}\right)^{-1} \ \ \ .
\end{equation}
Substituting this inequality into Eq.~(\ref{NPrat}) leads to an upper bound on the contribution from $\mathcal{O}_+$ to $\Delta\Gamma/\Gamma$ that decreases logarithmically as $\Lambda$ increases but grows quadratically with $m_h$. This bound is generally weaker than the expectation based on one-loop matching, but it will apply even in specific models that give a negligible renormalization of $\mathcal{O}_{eY}$. 

Before looking at the implications of the above naturalness constraints on the bounds for $\Delta \Gamma (h\to \tau^+ \tau^-)/\Gamma$, in the next section we explore possible constraints arising from the measurement of the $\tau$ anomalous magnetic moment.

\subsection{Anomalous magnetic moment constraints on $O_+$}

Since the coefficients $C_\pm$ of $\mathcal{O}_\pm$ depend on linear combinations of $C_{De}$ and $C_{{\bar D}e}$, one might expect $|C_+(\Lambda) |\sim |C_{-}(\Lambda)|$ in any NP scenario that gives rise to both operators. While $\mathcal{O}_{-}$ does not contribute to $h\to \ell^+\ell^-$ at tree-level, it does contribute to the $\ell$ anomalous magnetic moment. Specifically, 
$\mathcal{O}_{-}$ can be expressed in terms of the magnetic moment operators
\bea
\mathcal{O}_B & = & g_1 {\bar L} \phi \sigma^{\mu\nu} e\, B_{\mu\nu}  \\
\mathcal{O}_W & = & g_2 {\bar L}\tau^I \phi \sigma^{\mu\nu} e\, W^I_{\mu\nu} 
\eea
and $\mathcal{O}_{He}$ and $\mathcal{O}_{H\ell}^{(1)}$ by using the identity 
\begin{equation}
D^\mu D_\mu = {\not\!\! D}{\not\!\! D} +i\sigma_{\mu\nu} D^\mu D^\nu \ \ \ 
\end{equation}
and suitable integrations by parts, leading to
\bea
\label{eq:ominusrel}
O_{-} + {\rm h.c.} &=& - \frac{1}{4} \, \left(Y_L + Y_e \right) \, O_{B} - \frac{1}{4} \, O_{W} \nn \\
&-& y_e \, O_{He}  + y_e \,  O_{H\ell}^{(1)}  + {\rm h.c.}. 
\eea
After EWSB, one has
\bea
\mathcal{O}_B & \rightarrow & \frac{g_1 v}{\sqrt{2}}\,  {\bar \ell}\sigma_{\mu\nu} P_R\ell \left[-\sin\theta_W Z^{\mu\nu}+\cos\theta_W F^{\mu\nu}\right]\\
\nonumber
\mathcal{O}_W & \rightarrow &- \frac{g_2 v}{\sqrt{2}}\,  {\bar \ell}\sigma_{\mu\nu} P_R\ell \left[\cos\theta_W Z^{\mu\nu}+\sin\theta_W F^{\mu\nu}\right]\ \ \ ,
\eea
where $F^{\mu\nu}$ and $Z^{\mu\nu}$ are the field strength tensors for the $Z^0$ and photon respectively. Since $g_1\cos\theta_W=g_2\sin\theta_W=e$, we have
\begin{equation}
\frac{1}{\Lambda^2}\left[\mathcal{O}_B-\mathcal{O}_W +\textrm{h.c.}\right]\rightarrow 
\frac{\sqrt{2} e v}{\Lambda^2} {\bar\ell} \sigma_{\mu\nu} \ell\, F^{\mu\nu} \ \ \ .
\end{equation}
Using this result, together with Eq.~(\ref{eq:ominusrel}) and the definition of the anomalous magnetic moment $a_\ell$
\bea
a_\ell \equiv \frac{g_{\ell} - 2}{2},
\eea
as the coefficient of the operator
\bea
\frac{e}{4m_\ell}\, {\bar\ell} \sigma_{\mu\nu}\ell\, F^{\mu\nu}
\eea
we obtain 
\begin{equation}
\label{eq:anom1}
 \delta a_\ell[\mathcal{O}_{-}]  = 2 \, \sqrt{2} \,  y_\ell \, \left(\frac{v^2}{\Lambda^2}\right)\,  \, C_{-} . 
\end{equation}

The $\tau$ anomalous magnetic moment 
has never been directly measured.
The best bound is given by DELPHI \cite{Abdallah:2003xd}  which finds the $95 \%$ CL
\bea
-0.052 < a_\tau < 0.013.
\eea
The current standard model calculation \cite{Eidelman:2007sb} of $a_\tau$ is
\bea
 a_\tau = 117 721 (5) \times 10^{-8}
\eea
This leads to a conservative estimate of the deviation from the SM, considering the lack of data,
given by
\bea
\delta a_\tau < 1 \times 10^{-3}. 
\eea
Using this bound and Eq.~(\ref{eq:anom1}) leads to
\begin{equation}
\label{eq:anom2}
\left| C_{-} \right| \, \frac{v^2}{\Lambda^2} <  0.05\ \ \ .
\end{equation}

If  NP at high scales leads to
\bea
|C_{De}| \sim |C_{{\bar D} e}| \sim |C_{De}-C_{{\bar D}e}| \sim |C_{De}+C_{{\bar D}e}|,
\eea
then Eq.~(\ref{eq:anom2}) would imply an upper bound on the contribution from $\mathcal{O}_+$ to $\Delta\Gamma/\Gamma$. 

\begin{figure}[hbt]
\centerline{\scalebox{1.2}{\includegraphics{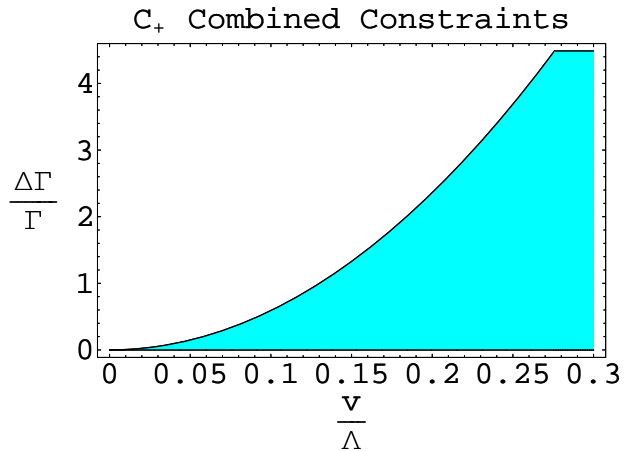}}}
\caption{The combined constraints on contributions from $\mathcal{O}_+$ due to: 1) the  anomalous magnetic moment constraints on $C_{+}$ in the 
NP scenario where $|C_{De}| \sim |C_{{\bar D} e}| \sim |C_{De}-C_{{\bar D}e}| \sim |C_{De}+C_{{\bar D}e}|$
and, 2) the naturalness constraint on $O_{+}$ due to the
one-loop matching contribution to $\mathcal{O}_{eY}$ from $\mathcal{O}_+$.
In both cases we have set $m_h =  140 \, {\rm GeV}$.}
\label{momnatcombined}
\end{figure}

In Fig.(\ref{momnatcombined}) we plot the bound on the contribution of $O_+$ to $\Delta \Gamma (h\to \tau \tau)/\Gamma$ as a function of the NP scale $\Lambda$ for the choice  $m_h=140$ GeV. The curved solid line comes from the naturalness bound in Eq.(\ref{eq:Cplusloop}) and the horizontal
solid line comes from the magnetic moment constraint of Eq.(\ref{eq:anom2}). As seen in this figure, the $m_\tau$-naturalness bounds dominate at higher values of the NP scale $\Lambda$. Only for lower values of the NP scale $\Lambda < 1$ TeV do the magnetic moment constraints become important for $h\to \tau^+ 
\tau^-$. It is also possible for NP above $\Lambda$ to dictate
$|C_+| \gg | C_-|$ since $O_+$ and $O_-$ are independent operators. In this case the constraints 
from $\delta a_i$ on $C_-$ will not apply to $C_+$ even for $\Lambda < 1 \, {\rm TeV}$.

\begin{figure}[hbt]
\centerline{\scalebox{1.1}{\includegraphics{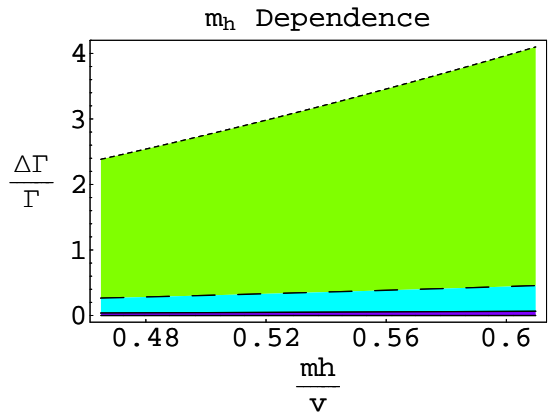}}}
\caption{Higgs mass dependence on the $O_{+}$ naturalness bound on $\Delta \Gamma/\Gamma$ for the range $114.4 \, {\rm GeV} < m_h < 150 \, {\rm GeV}$.
The dotted curve is the upper naturalness bound for the NP scale $\Lambda = 1 \, {\rm TeV}$,
the dashed curve for  $\Lambda = 3 \, {\rm TeV}$ and 
 the solid curve  for $\Lambda = 8 \, {\rm TeV}$. }
\label{mhdependence}
\end{figure}
It is also interesting to consider the Higgs mass dependence of the contribution of $O_+$ to $\Delta \Gamma/\Gamma$ for a fixed value of $\Lambda$. This is shown in Fig.(\ref{mhdependence})
for  $\Lambda  = 1,3,8 \,  {\rm TeV}$ where the naturalness constraints dominate over the anomalous magnetic moment constraints. We see that in general the naturalness bounds on $\Delta \Gamma /\Gamma$ become weaker for increasing Higgs mass values.

Although $h\to \mu^+ \mu^-$ is experimentally not a promising channel it is interesting to note that  $\Delta \Gamma(h\to \mu^+ \mu^-)/\Gamma$ could be as large as $20\%$ in spite of the extremely stringent constraint coming from the experimental bound on the muon anomalous magnetic moment.  The muon anomalous
magnetic moment has been measured very precisely
\cite{PDBook}: 
\bea
\delta a_\mu < 22(10) \times 10^{-10}.
\eea
The corresponding constraint in this case will be 
\begin{equation}
\left| C_{-}^\mu \right| \, \frac{v^2}{\Lambda^2} <  2 \times 10^{-4},
\end{equation}
from which we conclude a possible  $20\%$ effect in 
$\Delta \Gamma (h\to \mu^+ \mu ^-)/\Gamma$. This is due to an enhancement coming from two Yukawa factors. The first Yukawa factor appears in the standard way as shown in Eq.(\ref{NPrat}). An additional Yukawa factor appears as seen in Eq.(\ref{eq:anom1}) which dictates the bound on $C_-$ due to the anomalous magnetic moment constraint. These two enhancements give $\Delta \Gamma /\Gamma \sim 1/y_{\mu}^2$ allowing for a sizable effect. For this reason, realistic future improvements in the measurement of the $\tau$ anomalous magnetic moment are unlikely to severely constrain $\Delta \Gamma (h\to \tau^+ \tau ^-)/\Gamma$.


\section{Implications and Conclusions}
\label{NPeffects}

Without a specific model for NP whose low-energy effects are characterized by the $n=6$ effective operators, we cannot make precise quantitative predictions for $\Delta\Gamma/\Gamma$. It is, nevertheless, possible to identify four possible scenarios that could result from integrating out the massive degrees of freedom and estimate the size of their effects using naturalness criteria consistent with experimental constraints .  From the standpoint of $h\to\ell^+\ell^-$ decays, these scenarios can be described in terms of  the operators $\mathcal{O}_{eH}$ and $\mathcal{O}_{+}$ since whatever set of $n=6$ operators arises from integrating out the heavy physics can be related to $\mathcal{O}_{eH}$ and $\mathcal{O}_{+}$ by using the equations of motion.
The four scenarios and the corresponding expectations for $\Delta\Gamma/\Gamma$ are  as follows.
\begin{itemize}
\item[(i)] $C_{eH}(\Lambda)\not=0$, $C_{+}(\Lambda)=0$. An expected upper bound on $\Delta\Gamma/\Gamma$ is given by Fig. 2.
\item[(ii)] $C_{eH}(\Lambda)=0$, $C_{+}(\Lambda)\not=0$. An expected upper bound on $\Delta\Gamma/\Gamma$ is given by Figs. 5 and 6.
\item[(iii)] $C_{eH}(\Lambda)\not=0$ and $C_{+}(\Lambda)\not=0$. If $|C_{eH}(\Lambda)|\sim |C_{+}(\Lambda)|$, then one would expect the maximum possible deviation $\Delta\Gamma/\Gamma$ to be given by 
some combination of curves in Figs. 2, 5 for a specific higgs mass dependence given in Figs. 6. A conservative estimate is to take the minimum of the curves in Figs. 2 and 5, assuming $m_h$ is not too different from $v$. 
\item[(iv)] $C_{eH}(\Lambda)=0=C_{+}(\Lambda)$, leading to no deviation. 
\end{itemize}

We observe that, in all but scenario (iv), an order one shift in the $h\to\tau^+\tau^-$ rate is possible when the scale of NP is $\sim$ TeV. 

The LHC can look for a deviation from the SM rate of up to 10\%  or more in $\Gamma (h\to \tau ^+ \tau ^-)$. Any such deviation would not invalidate the Yukawa mechanism but would be consistent with NP at TeV scales in addition to the SM Higgs. Thus, it is necessary to disentangle TeV scale effects before drawing any conclusions on the Higgs mechanism for fermion mass generation. The naturalness considerations discussed above imply that a 10\% or larger deviation for the decay rate from the SM prediction would be associated with a NP mass scale $\lesssim 10 \, {\rm TeV}$.


It is also interesting to examine these conclusions in the minimal lepton flavor violation (MLFV) \cite{Dugan:1984qf,Chivukula:1987py,Hall:1990ac,D'Ambrosio:2002ex,Cirigliano:2005ck,Ali:1999we,Buras:2003jf,Branco:2006hz} approach, wherein the flavor structure of higher-dimension operators is determined by appropriate insertions of the lepton Yukawa matrix, $\bf{Y}$. In the simplest case, the flavor structure of $\mathcal{O}_{eH}$ and $\mathcal{O}_+$ would be aligned with $\bf{Y}$, so that after diagonalization, the operator coefficients $C_{eH}$ and $C_+$  would be flavor diagonal and proportional to $y_\ell$ for a given generation. In order to saturate the naturalness upper bound on $\Delta \Gamma/ \Gamma$ in MLFV, large Wilson coefficients $C_{MLFV}  \sim 2 \sqrt{2} \Lambda^2/v^2 \gg 1$ are required
as seen in Eq.(14).  However, even with such a large Wilson coefficient  $C_{MLFV} \, y_\tau \sim 0.2$
which is still perturbative. Thus, the upper bound on naturalness can be easily realized in MLFV.

Operators of the form $\mathcal{O}_{eH}$ and $\mathcal{O}_+$ could 
lead to flavor changing effects if the flavor structures of $O_{eH}$ and $O_+$  are not aligned with $\bf{Y}$ after diagonalization. In MLFV, for example, the relationship between the flavor diagonal and off diagonal contributions of these operators can be fixed with a choice of field content \cite{Cirigliano:2005ck}. The off diagonal flavor changing effects due to $\mathcal{O}_{eH}$ and $\mathcal{O}_+$ contribute at one loop to the flavor-changing decays $\tau^- \to \ell_j^- \, \gamma$, where $\ell_j = \mu, e$. From a straightforward dimensional analysis, we find that the naturalness expectations discussed above imply contributions to the decay branching ratios
$B_{\tau\to \ell_j\gamma}=\Gamma(\tau\to \ell_j\gamma)/\Gamma(\tau\to \ell_j{\bar\nu} \nu)$
 that are well below the present experimental limits. For example, the contribution from $\mathcal{O}_{eH}$ to $B_{\tau\to e\gamma}$ is roughly $10^{-8}\,  (v/m_H)^4\, (C_{eH}v^2/\Lambda^2)^2$, so that for $C_{eH} v^2/\Lambda^2\sim y_\tau$ as implied by the tree-level naturalness considerations and for $m_H\sim v$, one obtains a contribution to $B_{\tau\to e\gamma}$ of order $10^{-12}$, a result that is seven orders of magnitude smaller the experimental limit. Thus, naturalness considerations lead to considerably more stringent expectations for the dimension six operators than one would infer from these flavor-changing decays.

\begin{acknowledgments}
We thank B. Grinstein, J.Kile,  A. Manohar, and M. Wise for useful conversations and comments on the manuscript.
This work supported in part by the US  Department of Energy
under contracts DE-FG03-97ER40546 (MT), DE-FG02-05ER41361 (MRM), and DE-FG03-92ER40701 (SM); 
and by the NSF under grant PHY-PHY-0555674 (MR-M).
\end{acknowledgments}
\newpage

\bibliographystyle{h-physrev3.bst}
\bibliography{Higgs}

\end{document}